\documentstyle[twocolumn,aps,epsf,prb]{revtex}
\begin{document}
\draft

\twocolumn[
\hsize\textwidth\columnwidth\hsize\csname@twocolumnfalse\endcsname

\title{Quantum computation with quasiparticles of the Fractional Quantum 
Hall Effect}

\author{D.V. Averin and V.J. Goldman}

\address{Department of Physics and Astronomy, SUNY, Stony Brook, 
NY 11794-3800, U.S.A.} 

\date{\today}
\maketitle
\begin{abstract}

We propose an approach that enables implementation of anyonic quantum 
computation in systems of antidots in the two-dimensional electron 
liquid in the FQHE regime. The approach is based on the adiabatic 
transfer of FQHE quasiparticles in the antidot systems, and uses their 
fractional statistics to perform quantum logic. Advantages of our scheme 
over other semiconductor-based proposals of quantum computation include 
the energy gap in the FQHE liquid that suppresses decoherence, and the 
topological nature of quasiparticle statistics that makes it possible 
to entangle two quasiparticles without their direct dynamic interaction.  

\end{abstract}

\vspace*{5ex}

]

\section{Introduction}

``Topological'' quantum computation with anyons has been suggested 
as a way of implementing intrinsically fault-tolerant quantum computation 
\cite{b1,b2,b3,b4}. Intertwining of anyons, quasiparticles of 
two-dimensional 
electron system (2DES) with non-trivial exchange statistics, induces 
unitary transformations of the system wavefunction that depend only on 
the topological order of the underlying 2DES. These transformations 
can be used to perform quantum logic, the topological nature of which 
is expected to make it more robust against environmental decoherence. 
The aim of this work is to propose specific and experimentally feasible 
approach for implementation of basic elements of the anyonic quantum 
computation using adiabatic transport of the fractional quantum Hall 
effect (FQHE) quasiparticles in systems of quantum antidots \cite{b5}. 

An antidot is a small hole in the 2DES produced by electron depletion, 
which localizes FQHE quasiparticles at its boundary due to combined 
action of the magnetic field and the electric field created in the 
depleted region. If the antidot is sufficiently small, the energy 
spectrum of the antidot-bound quasiparticle states is discrete, with 
finite excitation energy $\Delta$. When $\Delta$ is larger than the 
temperature $T$, modulation of external gate voltage can be used to 
attract quasiparticles one by one to the antidot \cite{b5,b6}. In 
this regime, adiabatic transport of individual quasiparticles in the 
multi-antidot systems can be used to perform quantum logic, in close 
analogy to adiabatic transport of individual Cooper pairs in systems 
of small superconducting islands in the Coulomb blockade regime 
\cite{b7}. In what follows, we describe specific designs of such 
logic gates, and discuss parameters of the FQHE qubits and mechanisms 
of decoherence in antidot systems. 

\section{FQHE qubits and logic gates}

As in the Cooper-pair qubits \cite{b7,b8,b9}, information in the 
FQHE qubits can be encoded by the position of a quasiparticle in the 
system of two antidots. The {\em FQHE qubit} (Fig.\ 1) is then the 
double-antidot system gate-voltage tuned near the resonance, where 
the energy difference $\varepsilon$ between the quasiparticle states 
localized at the two antidots is small, $\varepsilon \ll \Delta$. 
At energies smaller than $\Delta$, dynamics of such double-antidot 
system is equivalent to the dynamics of a common two-state system 
(qubit). The quasiparticle states localized at the two antidots are the 
$|0\rangle$ and $|1\rangle$ states of the computational basis of 
this qubit. The gate electrodes of the structure can be designed to 
control separately the energy difference $\varepsilon$ and the 
tunnel coupling $\Omega$ of the resonant quasiparticle states. 

\begin{figure}
\setlength{\unitlength}{1.0in}
\begin{picture}(3.0,1.7) 
\put(0.07,0.07) {\epsfxsize=3.in \epsfbox{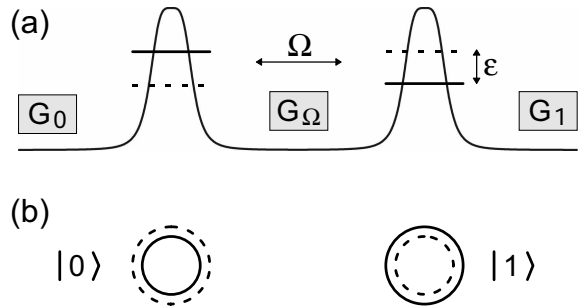}}
\end{picture}
\caption{Schematic energy profile (a) and structure (b) of the 
double-antidot FQHE qubit. Solid (dashed) lines (in (a), the   
horizontal lines) indicate the edges of the incompressible 
electron liquid when the quasiparticle is localized at the right 
(left) antidot. Displacement of the electron liquid is quantized 
due to quantization of the single-particle states circling the 
antidots. Dashed rectangles in (a) are the gate electrodes 
controlling the energies of the antidot states ($G_{0,1}$) and 
their tunnel coupling ($G_{\Omega}$).} 

\end{figure} 

The most natural approach to construction of the {\em two-qubit 
gates} with the FQHE qubits is to use fractional statistics 
\cite{b10,b11} of the FQHE quasiparticles. Due to this statistics, 
intertwining of the two quasiparticle trajectories in the course 
of time evolution of the two qubits realizes controlled-phase 
transformation with non-trivial value of the phase. Precise result 
of this operation depends on the nature of the FQHE state. In this 
work, we discuss the most basic and robust Laughlin state with 
the filling factor $\nu=1/m=1/3$, where the quasiparticles have abelian 
statistics and intertwining of trajectories leads to multiplication 
of the state wavefunction by the phase factor $e^{\pm 2\pi i/3}$. 
The sign of the phase depends on the direction of the magnetic field 
and the direction of rotation of one quasiparticle trajectory around 
another. 

A possible structure of the controlled-phase gate is shown in Fig.\ 2. 
Each of the columns of four antidots contains two qubits, and arrows 
denote trajectory of quasiparticle transfer through the system. 
The transfer leads to transformation of the quantum state of the two 
qubits and its shift from the gate input (left column in Fig.\ 2) 
to the output (right column). The quasiparticle 
transfer can be achieved by the standard adiabatic level-crossing 
dynamics. If a pair of antidots is coupled by the tunnel amplitude 
$\Omega$, a gate-voltage induced variation of the energy difference 
$\varepsilon$ through the value $\varepsilon=0$ (slow on the time 
scale $\Omega^{-1}$) leads to the transfer of a quasiparticle between 
these antidots. Correct operation of the controlled-phase gate in 
Fig.\ 2 requires that the gate voltage pulses applied to the antidots 
are timed so that the state of the upper qubit is propagated at 
first halfway through the gate, then the state of the lower qubit 
is propagated through the whole gate, and finally the state of the 
upper qubit is transferred to the output. In this case, if the 
quasiparticle of the upper qubit is in the state $|1\rangle$, 
trajectories of the quasiparticle propagation in the lower qubit 
encircle this quasiparticle, and the two states of the lower qubit 
acquire an additional phase difference $\pm 2\pi/3$, conditioned on 
the state of the upper qubit. We take the direction of magnetic field 
to be such that the state $|1\rangle$ of the lower qubit acquires 
a positive extra phase $2\pi/3$. Assuming that parameters of the 
driving pulses are adjusted in such a way that the dynamic phases 
accumulated by the qubit states are the multiple integers of 
$2\pi$, the transformation matrix $P$ of the gate can be written as  
\begin{equation}
P= \mbox{diag}\, [1,1,1,e^{2\pi i/3}]  
\label{1} \end{equation}  
in the basis of the four gate states $|00\rangle , \,  |01\rangle , 
\, |10\rangle , \, |11\rangle$.

\begin{figure} 
\setlength{\unitlength}{1.0in}
\begin{picture}(3.0,1.55) 
\put(0.05,0.07){\epsfxsize=3.in \epsfbox{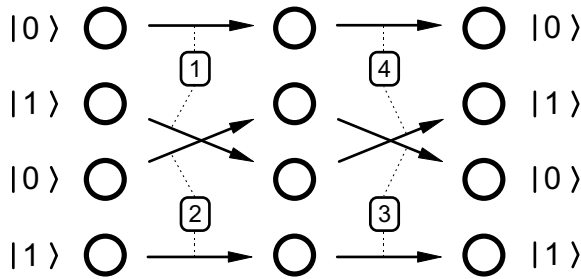}}
\end{picture}
\caption{A twelve-antidot two-quasiparticle implementation of the 
two-qubit controlled-phase gate. The states $|0\rangle$ and 
$|1\rangle$ are the computational basis states of the two qubits. 
The arrows show the quasiparticle transfer steps for each basis 
state during the gate operation. The arrow numbering denotes the 
time sequence of these steps. }
\end{figure} 

Controlled-phase gate (1) combined with the possibility of performing 
arbitrary singe-qubit transformations is sufficient for universal 
quantum computation. To demonstrate this explicitly, we construct 
a combination of the gate (1) with single-qubit gates that reproduces 
the usual controlled-NOT (C-NOT) gate $C$. The C-NOT gate is known to 
be sufficient for universal quantum computation \cite{b12}. Since the 
gates $P$ and $C$ are not equivalent with respect to single-qubit 
transformations, two applications of $P$ are required to reproduce 
$C$ \cite{b12a}. To find appropriate single-qubit transformations that 
should complement the two $P$'s, it is convenient to first reduce $P$ to 
the conditional $z$-rotation $R$ of the second qubit through angle $-\pi/3$, 
$R=\mbox{diag}\, [1,1,e^{-i \pi /3},e^{i\pi /3}]$. We notice that $R= 
S(-\pi/3)P$, where $S(\alpha)$ is an unconditional shift of the phase of 
the state $|1\rangle$ of the first qubit by $\alpha$. After this reduction, 
it is straightforward to find the necessary single-qubit transformations 
from the requirements that the state of the first qubit is unchanged 
by $C$, while the conditional action of $C$ on the second qubit is given 
by the Pauli matrix $\sigma_x$. These two requirements do not specify the 
necessary transformations uniquely. One possible choice is to use the 
transformations that correspond physically to modulation of the tunnel 
coupling between the states of the second qubit (i.e., involve only 
matrices $\sigma_x, \, \sigma_y$). In this case, we obtain
\begin{equation}
C=S(\pi/2)U_-^{\dagger}S(-\pi/3)PU_-U_+S(-\pi/3)PU_+^{\dagger} \, , 
\label{2} \end{equation} 
where $U_{\pm} = [\hat{1}]_1 \otimes [\exp \{ - i \varphi (\sigma_x \pm 
\sigma_y)/\sqrt{2} \} ]_2 $. Here the subscripts $1,2$ denote the part of 
the transformation acting on the first and the second qubit, respectively, 
and the rotation angle $\varphi$ is given by the condition $\cos 2\varphi 
=1/\sqrt{3} $, $\varphi \in [0,\, \pi/2]$. Physically, the transformations 
$S$ can be implemented as pulses of the gate voltage applied to the antidot 
$|1\rangle$ of the first qubit, while $U$s represent pulsed modulation of 
the amplitude of the tunnel coupling between the two antidots of the 
second qubit that keeps the phase of this coupling fixed.

\section{Decoherence mechanisms}

At low temperatures, the energy gap in the FQHE liquid exponentially 
suppresses quasiparticle excitations in the bulk of the sample. Due to 
this suppression, only sample edges and external metallic gate electrodes 
support low-energy excitations that can give rise to dissipation and 
decoherence in the antidot qubits. Qubit is coupled to both the gate 
electrodes and the edges by the Coulomb interaction. The charge $q$ of 
the qubit quasiparticle (for the primary Laughlin FQHE liquids, $q=e/m$, 
where $m$ is an odd integer) induces a polarization charge on the gate 
electrodes that oscillates in the course of qubit time evolution. The 
current induced in this way in the electrodes with finite 
resistance $R$ leads to energy dissipation and decoherence. This 
decoherence mechanism associated with ``electromagnetic environment'' 
of the structure (see, e.g., \cite{env}) is generic for most of the 
solid-state qubits. In the FQHE qubits, its strength should be lower 
than in other charge-based qubits, due to the smaller charge of the 
FQHE quasiparticles. Indeed, if the gate electrode is close to an 
antidot (on the scale of the distance $d$ between the two qubit antidots), 
the amplitude of the variations of the induced charge is roughly equal 
to the quasiparticle charge $q$. In this ``worst-case'' scenario, the 
limitation on the quality factor of qubit dynamics introduced by the 
gate electrode is equal to $e^2 R/\hbar m^2$ and is on the order 
of $10^{-3}$ for realistic values of the resistance $R$ and for $m=3$ 
qubits considered in this work. Optimization of the gate structure of 
the qubit should further reduce the strength of this type of decoherence 
by reducing electrostatic gate-qubit coupling.  

Coulomb interaction also couples qubit dynamics to edge excitations 
of the FQHE liquid. The edge supports one-dimensional (1D) chiral plasmon 
modes \cite{edge} propagating with velocity $v$. In the situation of 
interest here, when the qubit-edge distance $L$ is much larger than the 
qubit size $d$, the coupling operator $V$ can be expressed directly in 
terms of the 1D density $\rho (x)$ of charge carried by plasmon modes: 
$V=\sigma_z \int dx U(x)\rho (x)$. In this expression, $\sigma_z$ 
represents the position of the quasiparticle on one or the other antidot 
of the qubit, and $U(x)$ is the variation (with the quasiparticle 
position) of the electrostatic potential created by the qubit at point $x$ 
along the edge. A representative estimate of dissipation/decoherence rate 
introduced by this coupling is given by the decay rate $\Gamma$ of the 
excited antisymmetric superposition of the antidot states. Assuming 
that the qubit dipole is perpendicular to a straight edge, and that 
the electric field is not screened between the edge and the qubit, we can 
find $\Gamma$ directly: 
\begin{equation}
\Gamma = \left( \frac{d}{L} \right)^2 \left( \frac{e^2}{4\epsilon 
\epsilon_0 \hbar v} \right)^2 \frac{\Omega}{2\pi\hbar m^3} 
e^{-L\Omega/\hbar v} \, . 
\label{3} \end{equation} 
Here $\epsilon$ is the material dielectric constant. This equation 
shows that the edge-related limitation $\hbar \Gamma/\Omega$ on the qubit 
quality factor can vary widely depending on the system geometry and qubit 
energy parameters. For a realistic set of numbers, $\epsilon \simeq 10$, 
$v\simeq10^{5}$ m/s, $\Omega \simeq 0.1$K, $d\simeq 100$ nm (see the 
discussion below), we have $\hbar \Gamma/\Omega \simeq 10^{-3}$ for the 
edge that is $L\simeq 3$ $\mu$m away from the qubit. 

\section{Estimates and discussion} 

The basic set of conditions necessary for correct operation of the 
FQHE qubits and gates described above can be summarized as $T \ll 
\varepsilon, \, \Omega \ll \Delta$. The antidot excitation energy 
$\Delta$ is estimated as $\Delta\simeq \hbar u/r$, where $r$ is the 
antidot radius and $u\simeq 10^{4} \div 10^{5}$ m/s is the velocity 
of quasiparticle motion around the antidot \cite{b13}. This 
means that at a temperature $T\simeq 0.05$ K the radius $r$ should be 
smaller than 100 nm. Since the tunnel coupling $\Omega$ decreases rapidly
with the tunneling distance $s$ between the antidots, $\Omega \propto \exp 
\{ -eBs^2/12\hbar \}$ \cite{b14}, the fact that it should remain at least
larger than $T$ means that the distance between the tunnel-coupled antidots
should not exceed few magnetic lengths $l=(\hbar/eB)^{1/2} \simeq
10$ nm for typical values of the magnetic field $B$.  Although these
requirements on the radius $r$ and antidot spacing $s$ can be satisfied 
with the present-day fabrication technology, the necessity to control 
these parameters accurately presents a formidable challenge. It should 
be noted that this situation is not specific to our FQHE scheme, but 
characterizes all semiconductor solid-state qubits based directly on 
the quantum dynamics of individual quasiparticles, and not collective 
degrees of freedom (used, e.g., in the case of superconductors).

We believe that the challenges in fabrication of the FQHE qubits are 
well compensated for by the advantages of the FQHE approach. First 
of them is the energy 
gap of the FQHE liquid that suppresses quasiparticle excitations and 
associated decoherence in the bulk of the 2DES, and allows to control 
the remaining sources of decoherence through the system layout -- see the 
discussion above. The second advantage is the topological nature of 
statistical phase that makes it possible to entangle qubits without their 
direct dynamic interaction. This should lead to a simpler design of the 
FQHE quantum logic circuit in comparison to other solid-state qubits, 
where control of the qubit-qubit interaction typically presents a 
difficult problem.

\section*{Acknowledgments}

This work was supported by the NSA and ARDA under the ARO contract.

\end{document}